# Structural transformation for BaBiO$_{3-\delta}$ thin films grown on SrTiO$_3$-buffered Si(001) induced by an in-situ Molecular Beam Epitaxy cooldown process

I. Ahmed[a],[b],*, O. Richard[b], P. Carolan[b], M. Gambin[a],[b], L. Ceccon[a],[b], M. Kaviani[b], S. De Gendt[b],[c], C. Merckling[a],[b]

(a) Department of Materials Engineering, KU Leuven, Kasteelpark Arenberg 44, 3001 Leuven, Belgium.

(b) Imec, Kapeldreef 75, 3001 Leuven, Belgium.

(c) Department of Chemistry, KU Leuven, Celestijnenlaan 200F, 3001 Leuven, Belgium.

*: Corresponding author: islam.ahmed@imec.be

# Abstract

Oxygen loss is one of the common defect types in perovskite oxides whose formation can be caused by a low oxygen background pressure during growth or by straining the thin film. Crystalline $BaBiO_{3-\delta}$ thin films are grown by molecular beam epitaxy on $SrTiO_3$ buffered Si(001) substrates. Adsorption-controlled regime governs the epitaxy, as the sticking coefficient of bismuth is boosted by supplying activated oxygen at plasma power of 600 W during epitaxy. Even though activated oxygen is supplied during the growth process, large amount of oxygen vacancies is found to be created in the thin film depending on the cooldown process. Perovskite structure is obtained when the cooldown process includes an extended period of time during which activated oxygen at plasma power of 600 W is supplied. Another way for inducing the structural transformation is enabled via an ex-situ 600°C anneal step at molecular oxygen for the sample which includes oxygen vacancy channels. The transformation into perovskite structure $BaBiO_3$ is manifested as reconstructed octahedra based on transmission electron microscopy, Raman spectroscopy, and photoluminescence. Additionally, smaller out-of-plane lattice constant as well as increased monoclinicity are observed for the perovskite phase supported by X-ray diffraction data. In this paper, thermal mismatch and multivalency-facilitated tensile strain exerted on the layers by the underlying Si substrates are presented as the driving force behind the creation of oxygen vacancies.

I.   Introduction

Complex oxides have gained much research interest over the past few decades owing to their special properties which enable a variety of technologies. This includes but not limited to memory applications based on the ferroelectric $BaTiO_3$ and $PbTiO_3$, superconducting devices based on K-doped $BaBiO_3$ and $BaSbO_3$, and water splitting utilizing the photocatalytic activity of $NaTaO_3$ [1−4]. However, the challenging part for unveiling an improved device performance is lying within developing high quality material stacks with intrinsic properties [5]. Various techniques have been widely applied for growing perovskite oxides with nice quality such as physical vapor deposition (PVD), pulsed laser deposition (PLD), sputtering, and molecular beam epitaxy (MBE). High quality thin films with intrinsic properties have been developed by using conventional MBE, which is an important technique for the growth of perovskites with improved crystallinity and high purity. For instance, thin films of $BaSnO_3$ were reported to reach mobilities as high as those of the single crystals' only when grown by MBE [6].

MBE utilizes ultra-high vacuum environment to enable the long free path necessary for the molecular beams heading towards the substrate and for the electron beam required for in-situ reflection high energy electron diffraction (RHEED) [7]. Upon optimizing the co-deposition parameters, the epitaxial growth proceeds in a controlled fashion thanks to the lack of high energetic species. However, one of the widely known challenges for conventional MBE is the complete oxidation of certain cations [8]. Oxygen needs to be activated to form dissociated oxygen radicals which reach higher reactivity not possible with molecular oxygen. This is fundamentally a requirement for adsorption-controlled growth of $BaBiO_3$ where a volatile element such as bismuth has to be provided in large quantity in the presence of aggressive oxidizing agent [9].

Applying the suitable oxidant environment can tackle the issue of low sticking coefficient of bismuth, however, the formation of oxygen vacancies is still an issue because it also additionally depends on the residual strain energy built up within the thin film [10]. Loss of oxygen can transform the perovskite into a sub stoichiometric oxide, with different properties, which is common among complex oxides with a multivalent cation [11].

$SrCoO_{(3n-1)/n}$, $LaNiO_2$, and $LaCoO_{3-\delta}$ are examples of oxygen sponge materials which exhibit reversible intercalation/deintercalation of oxygen anionic planes [12−14]. Among common reasons for such occurrence of ordered oxygen vacancies are strain experienced by the thin films, growth in a reducing environment, or ionic liquid gating [15, 16]. According to a density functional theory (DFT) study, $SrFeO_{2.5}$ is obtained as oxygen vacancies order parallel to the surface under tensile strain, as this configuration maximizes the distance between the tetrahedral chains under that strain state [17]. As earlier reported, our $BaBiO_{3-\delta}$ thin films demonstrated a modulation of a $BiO_4$ tetrahedron every three $BiO_6$ octahedra due to the formation of ordered oxygen vacancy channels, resulting in an oxygen deficient phase [18]. These ordered oxygen vacancies are formed for the as-grown thin film with an MBE recipe described in our previous work with supplying activated oxygen radicals during the formation of the crystalline layer [9]. This structure possesses a mild breathing distortion, which is responsible for gapping the band structure [19].

In this context, the present work investigates the physical transformation occurring upon further oxidizing $BaBiO_{3-\delta}$ through different oxidative pathways. Cooldown at different activated oxygen conditions as well as post-annealing in molecular oxygen processes are checked for their capabilities in causing a transformation into a stoichiometric perovskite structure. The influence of the potential filling up of oxygen vacancies on the electronic structure, breathing distortion, and the optical response of the thin films is also studied. First-principles calculations were performed to gain better understanding of the perovskite matrix' oxygen loss pathway.

## II. Experimental

To investigate the effect of process control over the formation of ordered oxygen vacancies, $BaBiO_{3-\delta}$ thin films of 25 nm were epitaxially grown on Si(001) with a buffer layer of $SrTiO_3$(001). During growth, the temperature was fixed at 600°C while supplying activated oxygen species to the growth surface with a plasma power of 600 W, at a base chamber pressure around 1E-6 Torr. Bismuth was evaporated in higher quantities compared to barium to satisfy the adsorption-controlled growth conditions [8]. During the cooldown process, three samples had either molecular oxygen (0 W), activated oxygen (200 W), or activated oxygen (600 W) supplied for 5 minutes and the rest of the cooldown down process was carried out at cleaner vacuum. For another set of samples, the plasma power remained fixed at 600 W but the time during which the activated oxygen is supplied was extended to 10, 20, or 40 minutes instead of only 5 minutes, as elucidated in Fig. 1(a).

Additionally, sample A (cooled down in activated oxygen for 5 minutes with plasma power of 200 W) was ex-situ annealed in a molecular oxygen environment in an Annealsys AS-One 150 oven. Annealing was carried out at 600°C at 760 Torr. The ex-situ annealing in molecular oxygen is carried out to compare the anneal step with the 40-minutes cooldown process at plasma power of 600 W (sample B).

Growth mode and film morphology were monitored in real time by in-situ RHEED diagnostics with an electron gun operating at 20 kV. Crystalline structure of the thin films was assessed using a Phillips X'Pert Panalytical high resolution x-ray diffraction (HR-XRD) setup, equipped with Cu-K$_\alpha$ radiation of 1.54 Å wavelength.

Evaluating the full width at half-maximum (FWHM) of the rocking curve (RC) measurements allow for assessing the layers' crystalline quality. Reciprocal space maps (RSM) were also collected to gain better understanding for the crystal structure and the possible strain field experienced by the layers. The crystalline structure was also evaluated based on transmission electron microscopy (TEM) images. A Titan tool with a 200 kV operating voltage was used, with a very low electron beam intensity to minimize the material damage as much as possible. For quantifying the known breathing distortion in the thin films, Raman spectroscopy was utilized with a laser of 633 nm wavelength in backscattering configuration on a Horiba Jobin-Yvon LabRAM HR800 tool. Optical band gap was investigated by the photoluminescence (PL) responses which were collected at room temperature utilizing the same Raman tool. X-ray photoemission spectroscopy (XPS) for the samples is carried out using monochromatized Al $K_{\alpha 1}$ radiation with a beam energy of 1486.6 eV, in the angle integrated mode using a Physical Electronics QUANTES instrument.

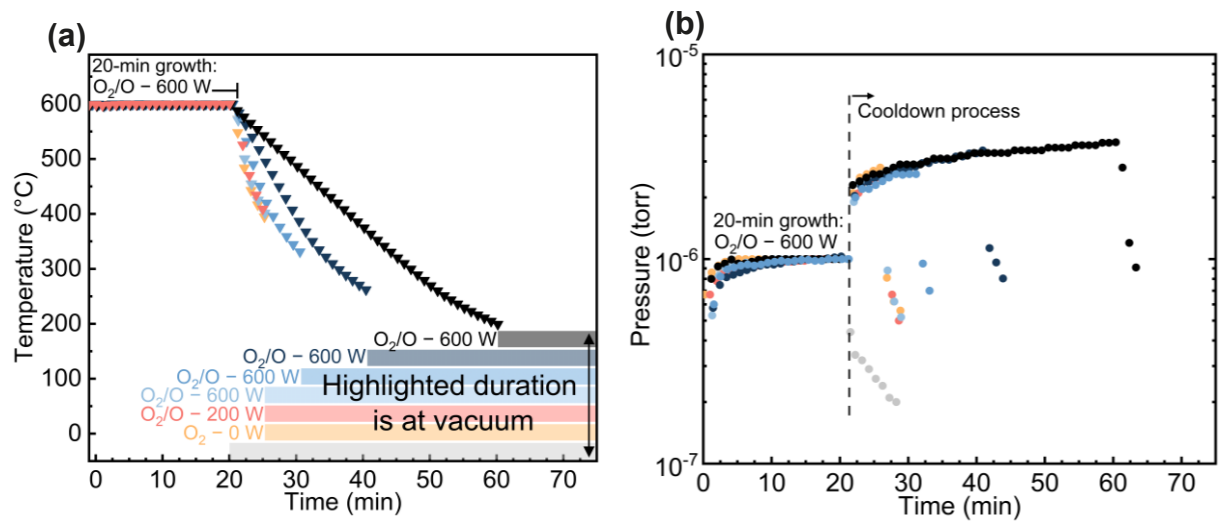

FIG. 1. Data retrieved from MBE tool's software showing (a) temperature in function of time for the growth process of 20 minutes and the cooldown process for the different samples at different oxygen conditions as stated next to the colored bars. Colored bars represent the time during which the cooldown is carried out in vacuum. (b) Pressure data as a function of time for both the growth and cooldown processes for the different samples.

DFT calculations were performed utilizing the widely known CP2K software packages, which take into account the mixing of Gaussian basis set with an auxiliary plane-wave basis set to solve the Kohn-Sham equations [20, 21]. The DZVP-MOLOPT-SR basis sets were employed on barium, bismuth, and oxygen atoms in conjunction with the GTH pseudopotential [22, 23]. The plane-wave cutoff was set to 900 Ry, which guarantees a convergence on the total energy. Additionally, to avoid the bond length overestimation typical for GGA functionals, the PBEsol exchange-correlation functional was used instead, which is known for resulting in lattice constants with a relatively higher accuracy compared to the PBE functional [24, 25]. $BaBiO_{3-\delta}$ cell was constructed from the fully relaxed structure as a $BiO_2$ terminated 40-atoms supercell.

## III. Results and discussions

The obtained layers are studied by XRD, and the results are shown in Fig. 2. Samples which were entirely cooled down in vacuum and partly in molecular oxygen have amorphous layers with no diffraction. For the 5-minutes cooldown process with activated oxygen at 200 W and 600 W, both layers have an out-of-plane lattice constant of 4.36 Å. Another interesting observation is that as the time during which the surface is provided with activated oxygen at 600 W increases the out-of-plane lattice constant shrinks, as clarified in Tab. 1. This effect is observed only for $BaBiO_{3-\delta}$ layers, while the out-of-plane lattice constant for $SrTiO_3$ remains unchanged at 3.91 Å.

Fig. 3(a) & (d) show the time evolution of RHEED patterns along the <100> crystallographic directions during the first 120 seconds of the growth with a step of 20 seconds as well as the RHEED patterns after cooling down for sample A (5 minutes / 200 W) and sample B (40 minutes / 600 W), respectively. It is noticed that the in-plane lattice constant does not evolve with time since the beginning of the growth as compared to that at the surface of the 25 nm thin films after cooling down. This confirms that both sample A and sample B have domain matching epitaxy as the mechanism for the growth of BaBiO$_{3-\delta}$ on SrTiO$_3$, given the large lattice mismatch ($f \approx$ 11.77%), where 8 unit cells of BaBiO$_{3-\delta}$ are accommodated on 9 unit cells of SrTiO$_3$ as illustrated in our previous TEM study [18]. Despite the difficulty to obtain images of low noise levels due to the fragility of this material upon e-beam exposure [26], TEM still gives important insights about the crystal structure. In the TEM images in Fig. 3(b) and Fig. 3(c), it is obvious that the cooldown for 5 minutes at 200 W of oxygen plasma results in a layer of BaBiO$_{3-\delta}$ with ordered oxygen vacancies, as observed based on the darker contrast areas aligned parallel to the substrate surface. Upon extending the time, during which activated oxygen is supplied at higher plasma power of 600 W, to 40 minutes, the oxygen vacancies are sufficiently filled up to retrieve the perovskite structure.

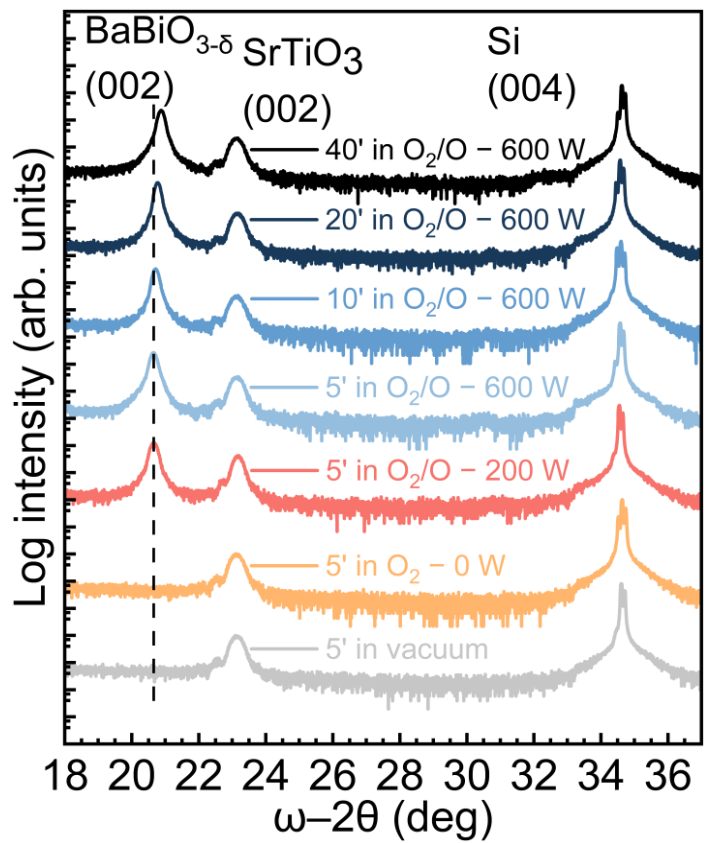

FIG. 2. XRD's out-of-plane symmetric scans showing diffraction from Si substrates' (004), SrTiO$_3$ buffer layers' (002), and BaBiO$_{3-\delta}$ active layers' (002) planes. XRD data highlights the effect of both the oxidative environment during cooldown process as well as the time during which activated oxygen is supplied at 600 W.

| Cooling time (min) | c (Å) |
|---|---|
| 5.0 | 4.36 |
| 10.0 | 4.35 |
| 20.0 | 4.34 |
| 40.0 | 4.32 |

TAB. 1. Lattice parameters of BaBiO$_{3-\delta}$ as a result of varying the cooldown time.

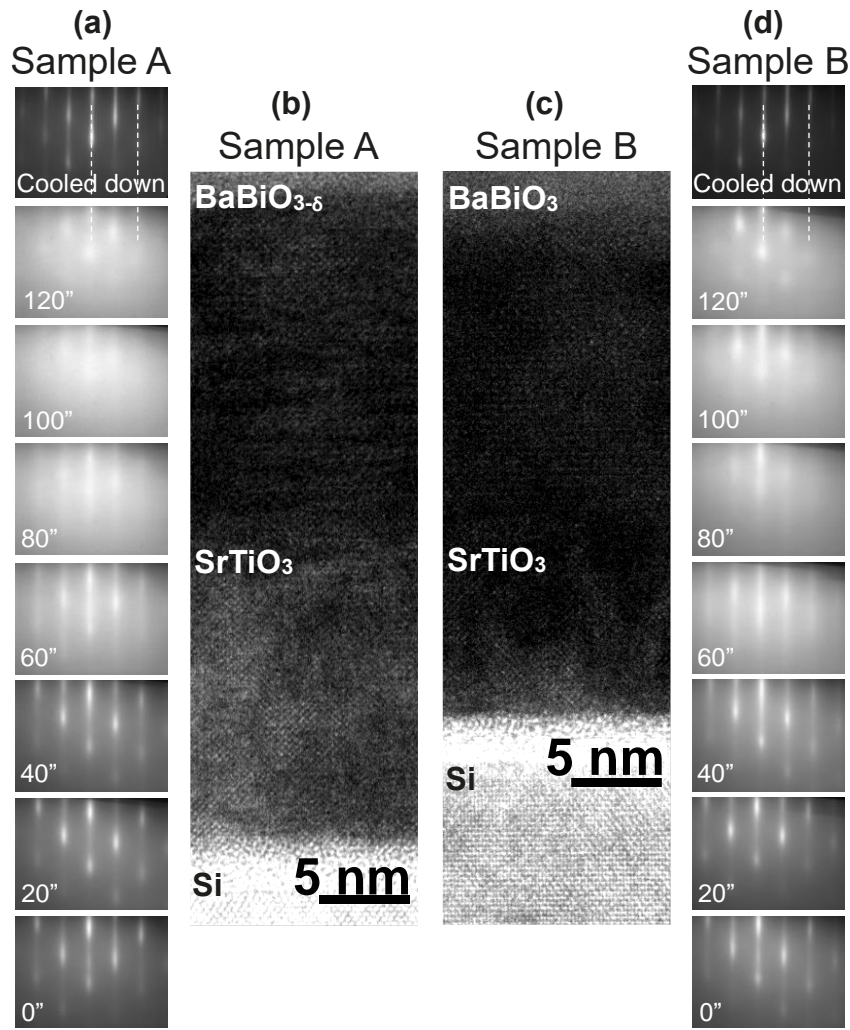

FIG. 3. (a) and (d) show the time evolution of the RHEED patterns along [100] crystallographic direction during the first 120 seconds of the growth as well as after cooling down of the thin film for sample A and sample B, respectively. White dashed lines are drawn vertically for the reference. Fig. 2(b) and (d) illustrate TEM images of the heterostructures: 25-nm-BaBiO$_{3-\delta}$/15-nm-SrTiO$_3$/Si(001) and 25-nm-BaBiO$_3$/10-nm-SrTiO$_3$/Si(001), respectively, with a 5 nm long scale bar.

Monoclinic charge ordering is a characteristic property for $BaBiO_3$ which results in the occurrence of breathing lattice distortion due to the double valency on the bismuth ionic sites ($Bi^{3+}$ & $Bi^{5+}$). The breathing distortion on the $BiO_6$ octahedra is manifested in the layer as a Raman spectral response at 565 cm$^{-1}$ [19]. According to Fig. 4(a), there is a correlation between the lattice constants of the layers and the Raman spectral intensity. Highest intensity response is recorded when activated oxygen is provided for 40 minutes during the cooldown process which results in the smallest out-of-plane lattice constant layer (perovskite structure). This distortion is fundamentally responsible for opening a Peierl's gap in the band structure of around 2.07 eV, according to a DFT study [27]. Therefore, the influence of the cooldown process on the PL is like that observed for the Raman spectra. The layers go from mild optical response at 2.19 eV when ordered oxygen vacancies exist in the structure to a more pronounced response when the perovskite structure is retrieved. While no response is recorded for the amorphous layers as expected.

Results presented so far are based on the in-situ cooldown process in the MBE chamber. However, ex-situ anneal process at molecular oxygen is also discussed in this study. Interestingly, as presented in Fig. S1(a), anneal process at molecular oxygen results in a reduction in the out-of-plane lattice constant of 1% for sample A (going from 4.36 Å down to 4.32 Å). Upon comparison with sample B, the ex-situ anneal for 5 minutes at 600°C in atmospheric pressure molecular oxygen environment has the same effect on the layer. The reduction in the out-of-plane lattice constant for $BaBiO_{3-\delta}$ is accompanied by approximately 30% increase in the layers' quality according to the FWHM of the RC as demonstrated in Fig. S1(b). Ex-situ anneal process also results in an enhancement in the breathing distortion and the optical response, according to Fig. S2(b) & (c), due to the transformation into the perovskite structure with higher octahedral content.

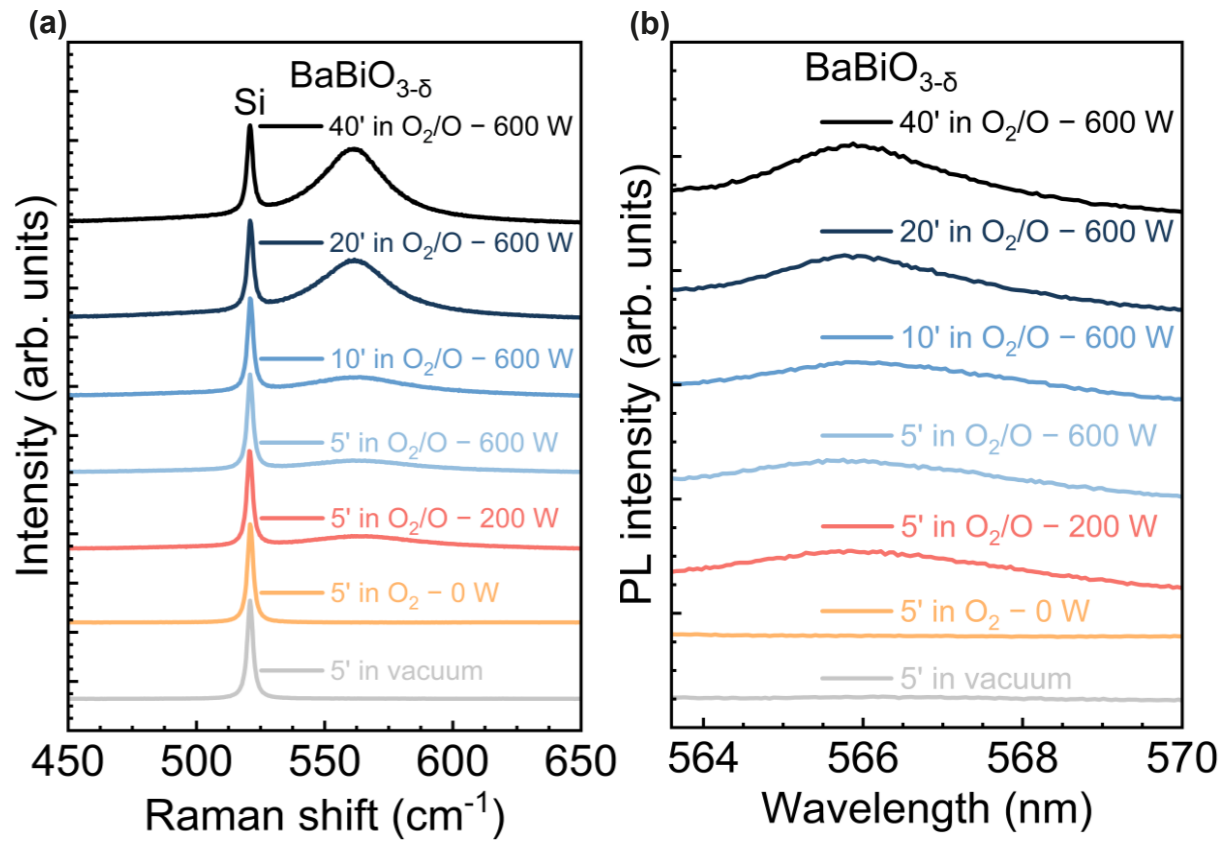

FIG. 4. (a) Raman spectra (λ = 633 nm) normalized based on Si active mode at 521 cm$^{-1}$, and (b) PL spectra (λ = 532 nm), showing the response for BaBiO$_{3-\delta}$ at 2.19 eV. Both measurements were carried out at room temperature for all the different samples.

To further evaluate the crystal structure, RSM's are collected for sample A, sample B, and sample A after ex-situ anneal process. The RSM's reported in Fig. 5 show the (103) reflections from BaBiO$_{3-\delta}$ and SrTiO$_3$ thin films. ($Q_x$ & $Q_z$) axes are normalized based on SrTiO$_3$ lattice parameters. The scattering vector components $Q_x$ and $Q_z$ were calculated and plotted using MATLAB, according to the geometric definition $Q = k_{out} - k_{in}$, where $k_{out}$ and $k_{in}$ are the diffracted and incident wavevectors, respectively. Additionally, the scattering vectors in terms of real lattice vectors for a monoclinic cell are represented as: $Q_x = \frac{k}{a}, Q_z = \frac{l}{c.\sin(\beta)}$. The information provided by the out-of-plane scattering vector ($Q_z$) component has contributions both from the out-of-plane lattice constant (c) and the angle between the c-axis and the ab-plane (β). Indeed, evaluating β is important for $BaBiO_{3-δ}$ since the material's monoclinicity is closely attributed to the spectroscopic data via the breathing distortion. To evaluate the two parameters separately, additional measurements were conducted and a mathematical procedure described in subsection C in supplementary material is developed. Based on the identified position of the $BaBiO_{3-δ}$'s (103) peaks in the RSM's, the in-plane, out-of-plane, and β parameters of the cells for the different samples are evaluated and tabulated in Tab. 2.

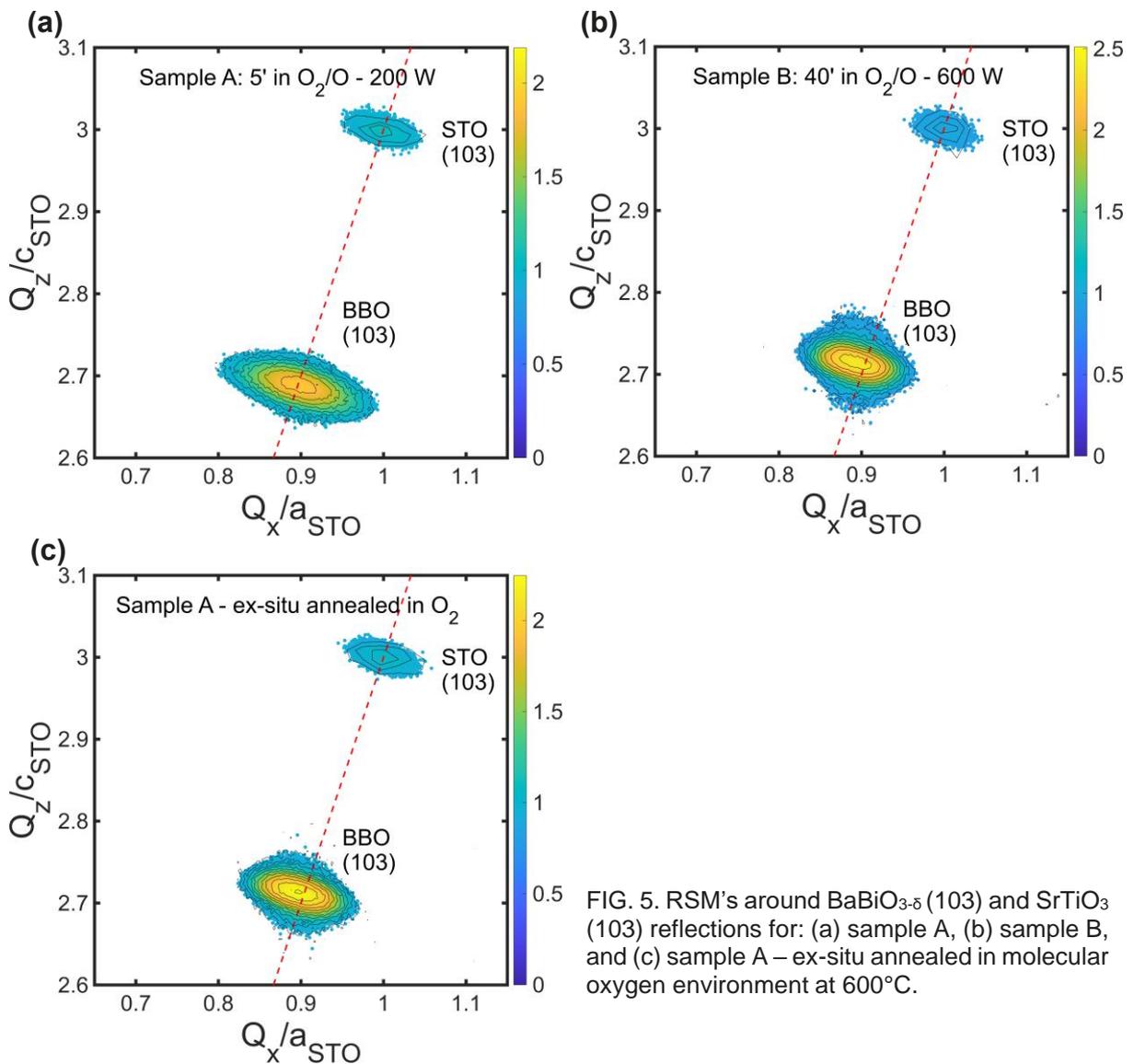

FIG. 5. RSM's around $BaBiO_{3-δ}$ (103) and $SrTiO_3$ (103) reflections for: (a) sample A, (b) sample B, and (c) sample A – ex-situ annealed in molecular oxygen environment at 600°C.

|  | a (Å) | b (Å) | c (Å) | β(°) |
|---|---|---|---|---|
| Sample A | 4.38 | 4.36 | 4.36 | 89.5 |
| Sample B | 4.40 | 4.36 | 4.32 | 93.3 |
| Sample A – annealed | 4.38 | 4.35 | 4.32 | 92.1 |

TAB. 2. Lattice parameters of $BaBiO_{3-\delta}$ based on the RSM's for different samples: sample A, sample B, and sample A after ex-situ anneal process.

For the three samples, the RSM's show totally relaxed $BaBiO_{3-\delta}$ layers with respect to $SrTiO_3$, confirming the presence of domain matching epitaxy. Sample A has lattice parameters closest to a cubic structure with $\beta = 89.5°$. The result of both the long cooldown process at 600 W and the ex-situ anneal is manifested as two effects: contraction in the out-of-plane lattice constant by 1% and an increased monoclinicity with $\beta = 93.3°$ & $92.1°$ for sample B & sample A after ex-situ anneal step, respectively.

Oxygen deficient $BaBiO_{3-\delta}$ is observed for the first time in thin film form, however it was observed in the past in bulk form and enabled an intensive topic of research, as researchers were investigating it as a candidate material for selective oxidation catalysis [28, 29]. Based on thermogravimetric experiments, at a fixed partial pressure and as a function of increasing temperatures, three distinct regions were established for $BaBiO_{3-\delta}$: $0 \leq \delta \leq 0.03$, $0.13 \leq \delta \leq 0.27$, and $0.42 \leq \delta \leq 0.45$, where expansion of the lattice constant upon oxygen loss was also observed [29]. The mechanism for the loss of oxygen for the thin films has not been investigated yet. However, it is possibly related to the thermal mismatch between the Si substrate and the $BaBiO_{3-\delta}$ thin film. A way to release the strain energy built up in the thin film upon cooling down rapidly could be via losing oxygen to the vacuum in an ordered fashion.

Due to the large lattice mismatch (11.77%), immediate misfit dislocation formation takes place at the earliest stage of epitaxy, which is confirmed by the delay in the RHEED streaky diffraction pattern formation, in Fig. 3(a) & (d), as the growth starts. For domain matching epitaxy, due to the large lattice mismatches, misfit strain does not play a major role in the accumulated residual strain. The grown thin films are fully relaxed at the high growth temperature sustaining their unstrained bulk lattice parameters [30]. As the thin film cools down, thermal strain has a major effective role in the relaxation process, in case there exists a thermal mismatch between the layer and the underlying substrate [31]. In our case, as Si has a thermal expansion coefficient of $2.6\times10^{-6}$ /°C, while that of $BaBiO_{3-\delta}$ is $1\times10^{-5}$ /°C (74% thermal mismatch), a tensile strain is built up within the thin film as it cools down from elevated temperature.

Tensile strain's relationship to formation of oxygen vacancies is known for perovskite oxides, even with little amount of tensile strain, the oxygen vacancies formation barrier is significantly lowered by reducing the efficiency of oxygen intercalation [32−34]. Additionally, tensile strain on $CaFeO_{2.5}$ stabilizes in the formation of ordered oxygen vacancies parallel to the substrate surface [10]. In another reduction reaction ($CaFeO_{2.5}$ to $CaFeO_2$), ordered oxygen vacancies are arranged perpendicularly regardless of the strain state of the thin film [35]. As reported in a DFT work, tensile strain not only results in increased bond length and octahedral rotation but also induces oxygen vacancies ordering because of the anisotropy of the energy of vacancy formation [36].

Stoichiometric $Ba_2Bi^{3+}Bi^{5+}O_6$ normally has equivalent amount of $Bi^{3+}$ and $Bi^{5+}$ which results in the monoclinic charge ordering. Studying the oxidation state of bismuth cations experimentally is important as it is the reason behind the monoclinic charge ordering. XPS measurements were carried out to have a closer look at the bismuth' core-levels for sample A, sample B, and sample A after ex-situ anneal process. Fig. S2(a) displays the core-level spectra for bismuth for the three different samples, fitted by the two doublets, while each doublet accounts for an electronic angular momentum: $4f_{5/2}$ and $4f_{7/2}$ (with a 5.3 eV shift). All samples possess $Bi^{3+}$ and $Bi^{5+}$ oxidation states with an energy separation between the two peaks of around 0.75 eV. However, with careful data fitting, it turns out that the ratio $Bi^{5+}/Bi^{3+}$ is different comparing one sample with another. $Bi^{5+}/Bi^{3+}$ goes from 0.95 for sample A to 1.17 after annealing the sample in high oxygen pressure. Indeed, this could be associated to the bismuth ions' environments being different according to the presence or absence of oxygen vacancy channels, which accordingly change the ionic coordination from tetrahedral to octahedral, respectively. Occurrence of oxygen vacancies triggers an imbalance of charge neutrality within the thin film, as each vacancy releases a charge of +2 ($V^{\bullet\bullet}_O$). Due to the multivalent nature of Bi ions, the thin film of $BaBiO_{3-\delta}$ can accommodate oxygen vacancies with high concentrations, even in an ordered fashion, via what is so called automatic charge neutralization. This strengthens the relationship between the strain state and the creation of ordered oxygen vacancies, similarly to perovskites containing transition metal ions [36].

As pointed out previously, according to the XRD data, this filling up of oxygen vacancies process is accompanied by a reduction in the out-of-plane lattice parameter and increased monoclinicity. This shrinkage is due to the reduced coulombic repulsion between the neighboring cations when oxygen vacancies are filled [37]. Additionally, $Bi^{3+}$ ions are oxidized into $Bi^{5+}$ ions which have smaller ionic radii ($Bi^{3+}-O \rightarrow 2.28$ Å & $Bi^{5+}-O \rightarrow 2.12$ Å) [38]. The increased monoclinicity is in place because as ordered oxygen vacancies are gotten rid of, the material retains its original structure which has an expected higher monoclinic charge ordering, as the octahedra are reconstructed.

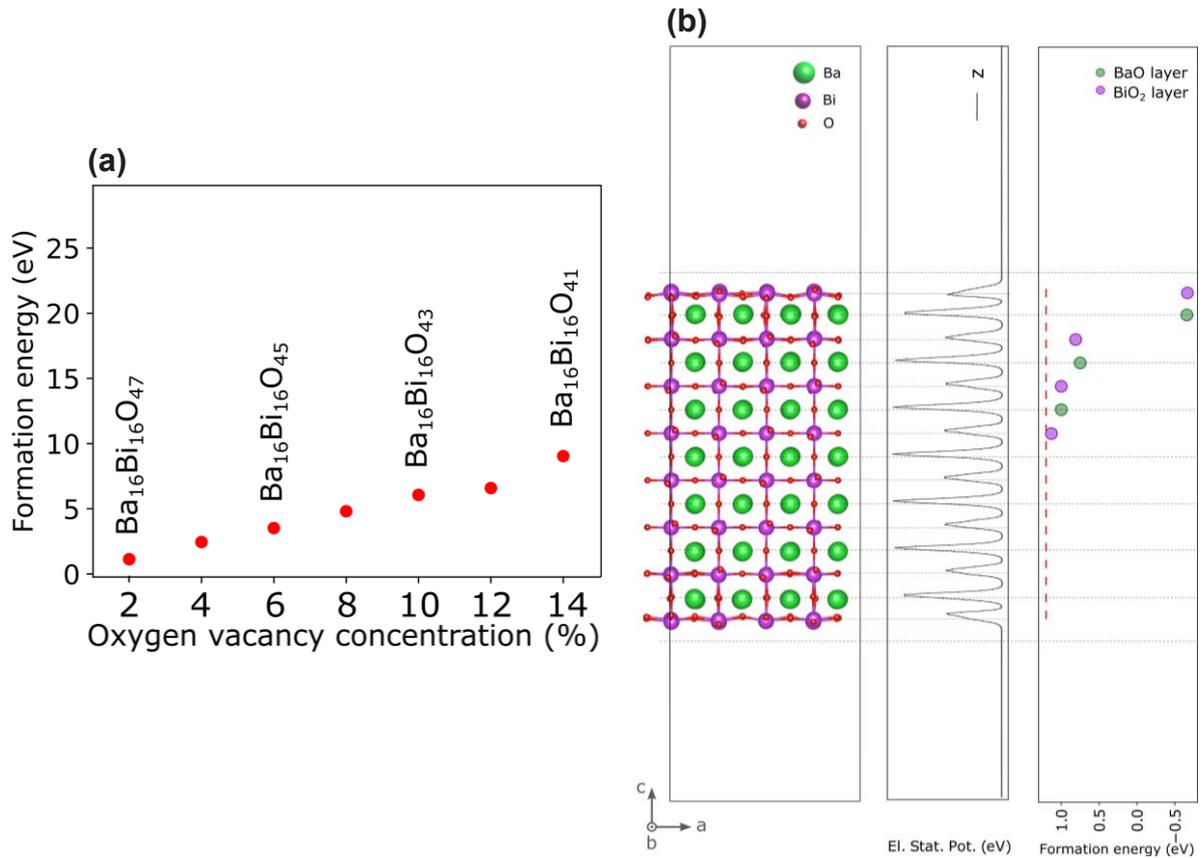

FIG. 6. DFT calculations showing (a) formation energies at different defect concentration for BaBiO$_3$ bulk, (b) evolution of the formation energies of V$_0$ at the surface of the slab compared to deeper into the film (red dashed line represents the formation energy of the bulk at this concentration).

Formation energies of the bulk oxygen vacancies at different concentrations ranging from 2% up to 14% in BaBiO$_3$ with c2/m crystal symmetry were evaluated by DFT calculations and presented in Fig. 6(a). Formation energies of oxygen vacancies in BaBiO$_3$ seem to be increasing as the defect concentration increases in the structure. An energy barrier of around 2.20 eV is calculated for BaBiO$_{3-\delta}$ with $\delta \approx 0.07$. This low energy of formation contributes alongside with the low oxidative environment at the MBE chamber to the spontaneous occurrence of ordered oxygen vacancies. Additionally, the formation energies of oxygen vacancies close to the terminated surface of a 40-atoms slab are compared to those deeper into the layer. In Fig. 6(b), it can be observed that the formation energy of a vacancy at the surface is lower than its formation deeper into the film (with energy barrier of around 1.5 eV). This indicates that oxygen vacancies are more likely to form at the surface region or migrate there if oxygen mobility is sufficiently high which drive the diffusion of the vacancies to the surface layer. This result leads to formulate a plausible pathway for the oxygen vacancies formation.

This pathway is described as follows: at such low oxidative environment during the cooldown process: Bi−O bonds are broken; oxygen ions diffuse to the surface and combine there as oxygen molecules which are released to the vacuum. This happens when the samples have a limited period during which activated oxygen is provided during the cooldown process. As observed in Fig. 1(b), the pressure built up during 5 minutes for sample A quickly drops to below the pressure attained during the growth (below 1E-6 torr). However, when the cooldown process includes an extended period during which activated oxygen is provided (sample B), the pressure built up is well maintained and even increases (up to 4E-6 torr) until the sample is cooled down already well below 200°C. This process seems to create an energy barrier for the formation of the oxygen vacancies within the film. This is, however, not the case for the sample which is cooled down entirely at vacuum or partly at molecular oxygen environment, that's the reason why the crystallinity is lost soon after the samples are completely cooled down.

## V. Conclusions

In summary, we throw light on the importance of the MBE in-situ cooldown process in regards of the oxygen vacancies formation in bismuthate perovskites. When limited duration of supplying activated oxygen during the cooldown process is used, thin film contains large amount of oxygen vacancies. As domain matching epitaxy is in place, the thermal strain largely constitutes the residual strain built up during the cooldown process, especially with the 74% thermal mismatch between BaBiO$_3$ and Si. The relaxation process for this tensile strain tends to take place in the form of oxygen loss in perovskite oxides, mainly in the presence of double valent cations, which is the case for BaBiO$_{3-\delta}$ with Bi$^{3+}$ & Bi$^{5+}$. Perovskite structure is still obtainable though by elongating the duration of supplying activated oxygen during the cooldown process or via ex-situ anneal step at atmospheric pressure environment of molecular oxygen. Structural transformation from oxygen deficient phase into perovskite is validated by XRD data which show a reduction in the out-of-plane lattice constant and increased monoclinicity, as expected for BaBiO$_3$. The progressive correlation between the decrease in the out-of-plane lattice constant and the increase in the octahedral content based on the measured breathing distortion Raman signature suggests a quantitative oxygen vacancy filling up via controlling the in-situ cooldown process. Fluorine-alloyed BaBiO$_{3-\delta}$ has gained research value as a material system because of the predicted topological insulating behavior, according to DFT, and the consequent development of Majorana-fermion based devices in case experimentally realized [39]. Regulating oxygen vacancy formation is a crucial step, as it will enable the incorporation of fluorine on the oxygen site in a well-controlled manner [40].

## Supplementary material

See the supplementary material which includes four subsections with relevant figures and equations. Subsections: "A: XRD data for the annealed sample", "B: XPS, Raman, and PL data for the annealed sample", and "C: Procedure for obtaining structural information for BaBiO$_{3-\delta}$".

## Acknowledgements

This work is part of IMEC's Industrial Affiliation Program. The authors would like to thank process and hardware engineers Hans Costermans and Kevin Dubois for their dedicated support on the MBE cluster tool. The authors prolong their gratitude to Stefanie Sergeant, Thomas Nuytten for Raman spectroscopy measurements and analysis and to Ilse Hoflijk, Thierry Conard for XPS measurements and analysis. This work has received funding from the European Research Council (ERC) under the European Union's Horizon 2020 research and innovation program (grant agreement No 864483).

## Data availability

The data that support the findings presented in this study can be made available by the corresponding author upon reasonable request.

## Author declarations

The authors have no conflicts to disclose.

# References

[1] Trithaveesak, O., J. Schubert, and Ch Buchal. "Ferroelectric properties of epitaxial BaTiO$_3$ thin films and heterostructures on different substrates." *Journal of applied Physics* 98.11 (2005).

[2] Venkatesan, Sriram, et al. "Monodomain strained ferroelectric PbTiO$_3$ thin films: Phase transition and critical thickness study." *Physical Review B* 78.10 (2008): 104112.

[3] Kim, Minu, et al. "Superconductivity in (Ba, K)SbO$_3$." *Nature Materials* 21.6 (2022): 627-633.

[4] Huerta-Flores, Ali M., et al. "Laser assisted chemical vapor deposition of nanostructured NaTaO$_3$ and SrTiO$_3$ thin films for efficient photocatalytic hydrogen evolution." *Fuel* 197 (2017): 174-185.

[5] Schlom, Darrell G. "Perspective: Oxide molecular-beam epitaxy rocks!." *APL materials* 3.6 (2015).

[6] Nunn, William, Tristan K. Truttmann, and Bharat Jalan. "A review of molecular-beam epitaxy of wide bandgap complex oxide semiconductors." *Journal of materials research* (2021): 1-19.

[7] Schlom, D. G., et al. "Oxide nano-engineering using MBE." *Materials Science and Engineering: B* 87.3 (2001): 282-291.

[8] Nunn, William, Tristan K. Truttmann, and Bharat Jalan. "A review of molecular-beam epitaxy of wide bandgap complex oxide semiconductors." *Journal of materials research* (2021): 1-19.

[9] Ahmed, I., S. De Gendt, and C. Merckling. "Self-regulating plasma-assisted growth of epitaxial BaBiO$_3$ thin-film on SrTiO$_3$-buffered Si (001) substrate." *Journal of Applied Physics* 132.22 (2022).

[10] Inoue, Satoru, et al. "Anisotropic oxygen diffusion at low temperature in perovskite-structure iron oxides." *Nature Chemistry* 2.3 (2010): 213-217.

[11] Khare, Amit, et al. "Topotactic metal–insulator transition in epitaxial SrFeO$_x$ thin films." *Advanced Materials* 29.37 (2017): 1606566.

[12] Karvonen, Lassi, et al. "The n = 3 member of the SrCoO$_{(3n-1)/n}$ series of layered oxygen-defect perovskites." *Materials Research Bulletin* 46.9 (2011): 1340-1345.

[13] Kawai, Masanori, et al. "Reversible changes of epitaxial thin films from perovskite LaNiO$_3$ to infinite-layer structure LaNiO$_2$." *Applied Physics Letters* 94.8 (2009).

[14] Biškup, Neven, et al. "Insulating ferromagnetic LaCoO$_{3-\delta}$ films: A phase induced by ordering of oxygen vacancies." *Physical Review Letters* 112.8 (2014): 087202.

[15] Petrie, Jonathan R., et al. "Enhancing perovskite electrocatalysis through strain tuning of the oxygen deficiency." *Journal of the American Chemical Society* 138.23 (2016): 7252-7255.

[16] Han, Hyeon, et al. "Control of oxygen vacancy ordering in brownmillerite thin films via ionic liquid gating." *ACS nano* 16.4 (2022): 6206-6214.

[17] Young, Joshua, and James M. Rondinelli. "Crystal structure and electronic properties of bulk and thin film brownmillerite oxides." *Physical Review B* 92.17 (2015): 174111.

[18] Ahmed, I., et al. "Influence of thickness scaling on the electronic structure and optical properties of oxygen deficient BaBiO$_{3-\delta}$ thin films grown on SrTiO$_3$-buffered Si (001) substrate." *APL Materials* 12.3 (2024).

[19] Lee, Han Gyeol, et al. "Anisotropic suppression of octahedral breathing distortion with the fully strained BaBiO$_3$/BaCeO$_3$ heterointerface." *APL Materials* 6.1 (2018).

[20] Hutter, Jürg, et al. "cp2k: atomistic simulations of condensed matter systems." *Wiley Interdisciplinary Reviews: Computational Molecular Science* 4.1 (2014): 15-25.

[21] Lippert, By Gerald, and JURG HUTTER and MICHELE PARRINELLO. "A hybrid Gaussian and plane wave density functional scheme." *Molecular Physics* 92.3 (1997): 477-488.

[22] VandeVondele, Joost, and Jürg Hutter. "Gaussian basis sets for accurate calculations on molecular systems in gas and condensed phases." *The Journal of chemical physics* 127.11 (2007).

[23] Goedecker, Stefan, Michael Teter, and Jürg Hutter. "Separable dual-space Gaussian pseudopotentials." *Physical Review B* 54.3 (1996): 1703.

[24] Perdew, John P., et al. "Restoring the density-gradient expansion for exchange in solids and surfaces." *Physical review letters* 100.13 (2008): 136406.

[25] Perdew, John P., Kieron Burke, and Matthias Ernzerhof. "Generalized gradient approximation made simple." *Physical review letters* 77.18 (1996): 3865.

[26] Harris, D. T., et al. "Charge density wave modulation in superconducting BaPbO$_3$/BaBiO$_3$ superlattices." *Physical Review B* 101.6 (2020): 064509.

[27] Franchini, Cesare, et al. "Structural, vibrational, and quasiparticle properties of the Peierls semiconductor BaBiO$_3$: A hybrid functional and self-consistent GW+ vertex-corrections study." *Physical Review B—Condensed Matter and Materials Physics* 81.8 (2010): 085213.

[28] Lightfoot, P., et al. "BaBiO$_{2.5}$, a new bismuth oxide with a layered structure." *Journal of Solid State Chemistry* 92.2 (1991): 473-479.

[29] Beyerlein, R. A., A. J. Jacobson, and L. N. Yacullo. "Preparation and characterization of oxygen deficient perovskites, BaBiO$_{3-x}$." *Materials research bulletin* 20.8 (1985): 877-886.

[30] Delhaye, G., et al. "Structural properties of epitaxial SrTiO$_3$ thin films grown by molecular beam epitaxy on Si (001)." *Journal of Applied Physics* 100.12 (2006).

[31] Rasic, Daniel, et al. "Structure-property correlations in thermally processed epitaxial LSMO films." *Acta Materialia* 163 (2019): 189-198.

[32] Hu, Songbai, et al. "Strain-enhanced oxygen dynamics and redox reversibility in topotactic SrCoO$_{3-\delta}$ (0< δ≤ 0.5)." *Chemistry of Materials* 29.2 (2017): 708-717.

[33] Petrie, Jonathan R., et al. "Strain control of oxygen vacancies in epitaxial strontium cobaltite films." *Advanced Functional Materials* 26.10 (2016): 1564-1570.

[34] Inkinen, Sampo, Lide Yao, and Sebastiaan van Dijken. "Reversible thermal strain control of oxygen vacancy ordering in an epitaxial La$_{0.5}$Sr$_{0.5}$CoO$_{3-\delta}$ film." *Physical Review Materials* 4.4 (2020): 046002.

[35] Inoue, Satoru, et al. "Anisotropic oxygen diffusion at low temperature in perovskite-structure iron oxides." *Nature Chemistry* 2.3 (2010): 213-217.

[36] Aschauer, Ulrich, et al. "Strain-controlled oxygen vacancy formation and ordering in CaMnO$_3$." *Physical Review B* 88.5 (2013): 054111.

[37] Ullmann, Helmut, and Nikolai Trofimenko. "Estimation of effective ionic radii in highly defective perovskite-type oxides from experimental data." *Journal of alloys and compounds* 316.1-2 (2001): 153-158.

[38] Cox, D. E., and A. W. Sleight. "Crystal structure of Ba$_2$Bi$^{3+}$Bi$^{5+}$O$_6$." *Solid State Communications* 19.10 (1976): 969-973.

[39] Yan, Binghai, Martin Jansen, and Claudia Felser. "A large-energy-gap oxide topological insulator based on the superconductor $BaBiO_3$." *Nature Physics* 9.11 (2013): 709-711.

[40] Yang, Jie, et al. "The role of F-doping and oxygen vacancies on the superconductivity in SmFeAsO compounds." *Superconductor Science and Technology* 22.2 (2008): 025004.

# Supplementary material

## A. XRD data for the annealed sample

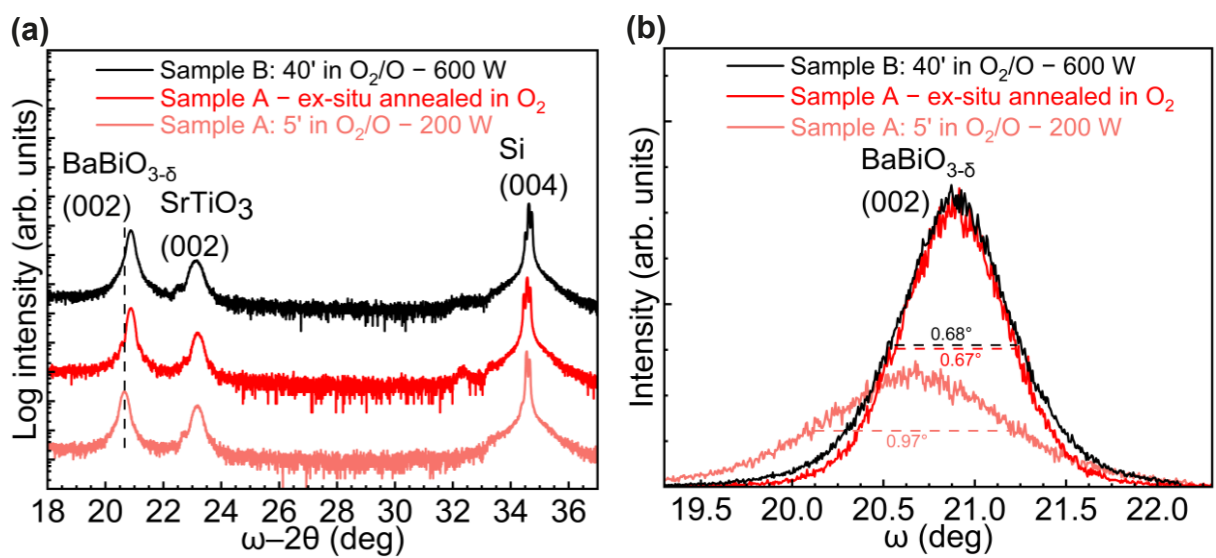

S1. (a) Out-of-plane XRD scans for the samples: sample A, sample B, and sample A after ex-situ anneal process in molecular oxygen environment, (b) Omega scans with color coded horizontal dashed lines highlighting the FWHM for each RC measurement (FWHM info: Sample A [$\Delta\omega(BaBiO_{3-\delta})$ = 0.97°], Sample B [$\Delta\omega(BaBiO_3)$ = 0.68°], Sample A – annealed in molecular oxygen at 600°C for 5 minutes [$\Delta\omega(BaBiO_3)$ = 0.67°]).

## B. XPS, Raman, and PL data for the annealed sample

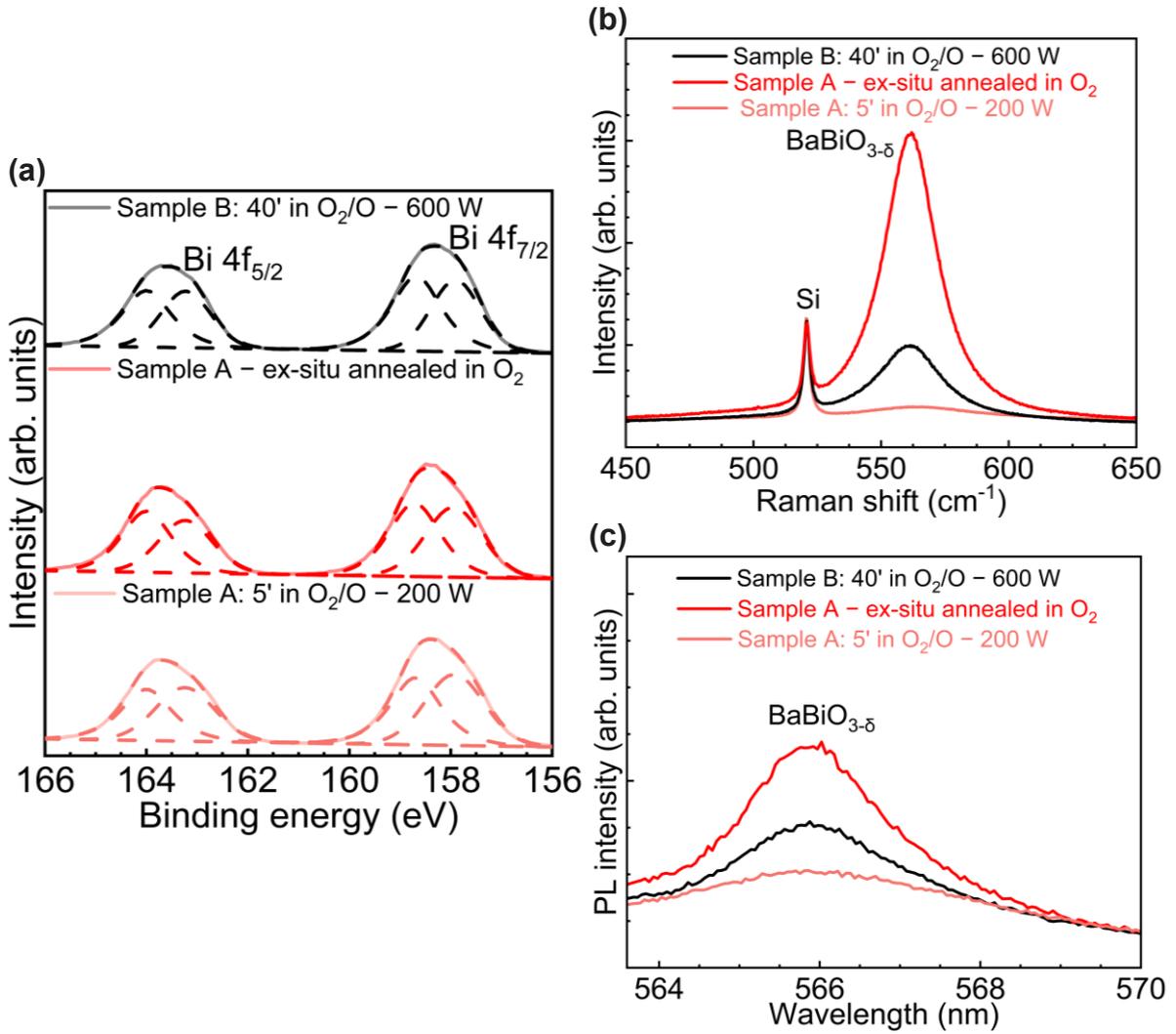

S2. (a) Bi 4f core-level spectra as measured by XPS for: sample A, sample B, and sample A after ex-situ anneal in molecular oxygen. (b) Raman spectra, (c) PL spectra, showing the optical response for BaBiO$_{3-\delta}$ at 2.19 eV, for the different samples.

## C. Procedure for obtaining structural information for BaBiO$_{3-\delta}$

To obtain the lattice parameters of the BaBiO$_{3-\delta}$'s unit cells, a total of five different reflections were measured. The (101), (011), and (002) peaks were acquired using ω−2θ symmetric scans. While the (222) and (103) reflections were captured performing asymmetric RSM's. This allows to adopt, as secondary scan axes, the $\chi$ and ω angular coordinates, respectively. Overall, the information extracted from the measurements are five times 2θ angular positions belonging to the five different BaBiO$_{3-\delta}$ peaks, as well as an additional $\chi$ coordinate provided by the RSM scan along the (222) direction. Having a total of six angular coordinates measured, equations were then set up using the lattice spacing formulas and a geometric relationship linking the $\chi$ coordinate of the (222) peak to the lattice parameters of the monoclinic cell as follows:

- (101) reflection:

$$\frac{1}{d_{101}^2} = \frac{1}{a^2 \sin^2(\beta)} + \frac{1}{c^2 \sin^2(\beta)} - \frac{2 \cos(\beta)}{ac \sin^2(\beta)}$$

- (011) reflection:

$$\frac{1}{d_{011}^2} = \frac{1}{b^2} + \frac{1}{c^2 \sin^2(\beta)}$$

- (002) reflection:

$$\frac{1}{d_{002}^2} = \frac{4}{c^2 \sin^2(\beta)}$$

- (222) reflection:

$$\frac{1}{d_{222}^2} = \frac{4}{a^2 \sin^2(\beta)} + \frac{4}{b^2} + \frac{4}{c^2 \sin^2(\beta)} - \frac{8 \cos(\beta)}{ac \sin^2(\beta)}$$

$$\chi(c, \beta) = \arctan\left(\frac{\| a(c) \cdot \hat{\imath} + (b(c) + c \cos(\beta)) \cdot \hat{\jmath} \|}{\| c \sin(\beta) \cdot \hat{k} \|}\right)$$

- (103) reflection:

$$\frac{1}{d_{103}^2} = \frac{1}{a^2 \sin^2(\beta)} + \frac{9}{c^2 \sin^2(\beta)}$$

The above-mentioned formulas were used in Wolfram Mathematica to form two systems of three equations that were solved independently. Namely, a, b, c and β to be determined, as each system's solution provides a function for three parameters as a function in the fourth. In an ideal scenario, the two systems will converge, however, due to inevitable measurement errors, the solution is varied. The overlapping of the solutions was imposed using the integrated Wolfram Mathematica optimization function *NMinimize*, allowing the angular coordinates to vary in a narrow interval around the initial estimated position. This procedure successfully leads to the identification of a unique set of angular positions satisfying all the equations and fitting the XRD patterns very well. These unique sets of angular coordinates were then used to calculate the lattice parameters of the unit cells of the different samples.